# A Semiparametric Joint Model for Terminal Trend of Quality of Life and Survival in Palliative Care Research

Zhigang Li[1*], H. R. Frost[1*], Tor D. Tosteson[1], Lihui Zhao[2], Lei Liu[2], Kathleen Lyons[3], Huaihou Chen[4], Bernard Cole[5], David Currow[6] and Marie Bakitas[7]

[1] Department of Biomedical Data Science, Geisel School of Medicine, Dartmouth College, Hanover, NH 03755

[2] Department of Preventive Medicine, Northwestern University, Chicago, IL 60611

[3] Department of Psychiatry, Geisel School of Medicine, Dartmouth College, Hanover, NH 03755

[4] Department of Biostatistics, University of Florida, Gainesville, FL, 32611

[5] Department of Mathematics and Statistics, University of Vermont, Burlington, VT, 05405

[6] Discipline of Palliative and Supportive Services, Flinders University, Bedford Park SA 5042, Australia

[7] School of Nursing, The University of Alabama at Birmingham, Birmingham, AL 35233

[*] These authors contributed equally to the paper

Corresponding: Zhigang Li, email: Zhigang.Li@dartmouth.edu

# A Semiparametric Joint Model for Terminal Trend of Quality of Life and Survival in Palliative Care Research

## ABSTRACT

Palliative medicine is an interdisciplinary specialty focusing on improving quality of life (QOL) for patients with serious illness and their families. Palliative care programs are available or under development at over 80% of large US hospitals (300+ beds). Palliative care clinical trials present unique analytic challenges relative to evaluating the palliative care treatment efficacy which is to improve patients' diminishing QOL as disease progresses towards end of life (EOL). A unique feature of palliative care clinical trials is that patients will experience decreasing QOL during the trial despite potentially beneficial treatment. Often longitudinal QOL and survival data are highly correlated which, in the face of censoring, makes it challenging to properly analyze and interpret longitudinal QOL trajectory. To address these issues, we propose a novel semiparametric statistical approach to jointly model longitudinal QOL and survival data. There are two sub-models in our approach: a semiparametric mixed effects model for longitudinal QOL and a Cox model for survival. We use regression splines method to estimate the nonparametric curves and AIC to select knots. We assess the model through simulation and application to establish a novel modeling approach that could be applied in future palliative care treatment research trials.

# 1. INTRODUCTION

Palliative care is a relatively new interdisciplinary specialty (medicine, nursing, social work, chaplaincy, and other specialties when appropriate) that focuses on improving quality of life for patients of any age with any serious disease, whether it can be cured or controlled, and for their families (Kelley and Morrison 2015). Data from 2010 have shown that palliative care programs were available or under development at more than 80% of large US hospitals (300+ beds) (Kelley and Meier 2010). Understanding of terminal trends or trajectories of QOL through carefully designed palliative care studies is crucial to help practitioners, patients and their care givers gain a better understanding of ill trajectories which may then help them feel in better control and empower them to cope with difficulties (Murray et al. 2005). Palliative care clinical trials present unique analytic challenges relative to evaluating the palliative care treatment efficacy in terms of improving patients' diminishing QOL as disease progresses towards EOL. QOL, symptom burden, anxiety, depression, and resource use, rather than survival, are common endpoints chosen to evaluate treatment efficacy of palliative care (Bakitas et al. 2009, Temel et al. 2010, Bakitas et al. 2015). These measures are captured longitudinally over relatively short time frames as patients' prognosis, especially in cancer palliative care trials, averages about a year. Additionally some patients may drop out or live beyond study end, resulting in longitudinal data and death times being censored. When analyzing palliative care studies, the interplay between longitudinal QOL and survival outcomes should be taken into account because these two outcomes typically strongly related, especially during EOL period, such that shortened survival and worsening QOL and symptoms are highly correlated. Without appropriate modeling of the correlation, estimates could be inefficient and potentially biased, and the clinical interpretation could be distorted. Censoring of the survival times occurs frequently due to dropout or study end, making it challenging to account for the correlation. To have the most accurate assessment of treatment efficacy (clinical effect) and avoid distorted interpretations and inefficient analyses due to censoring, it is important to thoughtfully consider survival-longitudinal data correlation in data analysis. To address these issues, we propose a novel statistical approach for jointly modeling longitudinal and survival data in palliative care trials.

In the statistical literature, significant effort has been devoted to methods for jointly modeling longitudinal and time-to-event data. Due to the complex nature of the joint distribution of longitudinal measurements $Y$ and time-to-event outcome $D$, two sub-models are commonly employed for the two types of outcomes, and these sub-models account for their dependency through conditional distribution or shared (or correlated) random effects. Comprehensive reviews on the differences and connections of these approaches can be found in the literature (Hogan and Laird 1997b, Tsiatis and Davidian 2004, Kurland et al. 2009, Rizopoulos and Lesaffre 2014) . One common approach, called shared parameter models, construct models for the two marginal distributions $f(Y)$ and $f(D)$ rather than factoring the joint distribution and accounting for the dependence between $Y$ and $D$ using shared or correlated random effects (Wulfsohn and Tsiatis 1997, Henderson, Diggle, and Dobson 2000, Elashoff, Li, and Li 2007, Liu, Wolfe, and Kalbfleisch 2007, Li et al. 2010, Barrett et al. 2015). Another general model, pattern-mixture models (Wu and Bailey 1989, Hogan and Laird 1997a, Fitzmaurice and Laird 2000, Su and Hogan 2010), factors the joint distribution $f(Y, D)$ into $f(Y|D)f(D)$ and models $f(Y|D)$ and $f(D)$ with two sub-models, where the sub-model for $f(Y|D)$ is usually modeled by a multivariate normal distribution conditional on the death time and hence the marginal distribution of $Y$ is a mixture of normal distributions over the survival distribution. In these approaches, the trajectories of longitudinal outcomes are usually assumed to have different patterns given different death or dropout times. A third general approach, the selection model (Diggle and Kenward 1994), factors the joint distribution in an opposite way by writing the joint distribution as $f(D|Y)f(Y)$, where the primary endpoint is usually survival. For longitudinal data prior to death, Kurland and Heagerty (2005) proposed a "partly conditional" approach that models the longitudinal outcomes conditional on the subjects being alive using weighted GEE where the survival data does not have to be explicitly modeled.

Another approach for longitudinal data prior to death, which we refer to as terminal trend model in this paper, is to jointly model the terminal trend of longitudinal data and survival by looking at the longitudinal data on the retrospective time scale from death time. Kurland et al. (2009) described a terminal trend model for analyzing only the decedents whose death times are observed. Unfortunately, the decedent analysis can generate substantial efficiency loss by omitting censored subjects when censoring rate is high (e.g., 47.34% for the motivating example described in Section 2). A decedent analysis could also generate biased results if the subgroup of decedents involves selection bias (Bach, Schrag, and Begg 2004). To alleviate the loss of power due to removing the censored subjects entirely, Li et al. (2013) developed a parametric terminal trend model specifically for cancer palliative care studies by appropriately accounting for censoring and this method has been applied in Bakitas (2015). In a related method designed for analyzing longitudinal cost data, Liu et al. (2007) proposed a 'turn back time' method for censored observations to characterize costs on a retrospective time scale which re-censors the longitudinal outcomes at some fixed time before the actual censoring time, while keeping the actual censoring time for the survival data.

The prominent advantage of a terminal trend model is that it directly estimates the terminal trend until death by modeling the longitudinal data on a retrospective time scale. Understanding the terminal trend is extremely important to help clinicians plan care to meet their patient's multidimensional needs better, and help patients and care givers cope with their situation (Murray et al. 2005). Therefore, terminal trend models are of particular interest for palliative care studies. The terminal trend model developed in Li et al. (2013) was a fully parametric model specifically proposed for studies with cancer patients. In this paper, we will develop a semiparametric terminal trend model with a semiparametric mixed effects sub-model for the longitudinal QOL outcomes and a Cox sub-model for the survival outcome. Use of nonparametric functions for the longitudinal trajectories allows us to build a much more flexible model that is not subjective to model misspecification and thus makes the model applicable to other disease populations as well (e.g., heart failure, dementia). Regression spline methods with natural cubic B-splines will be employed to estimate the unspecified nonparametric terminal trend. Quality-adjusted life years (QALY) can be also derived and analyzed using this terminal trend model.

The rest of this paper is organized as follows. Section 2 introduces the palliative care ENABLE III (Educate, Nurture, Advise, Before Life Ends, III) randomized clinical trial (RCT) that motivated this development. Section 3.1 gives notation and specifies the proposed statistical models. Estimation of the model parameters is detailed Section 3.2. The procedure for selecting the knots is described in Section 3.3. Estimation of QALY is presented in Section 3.4. A simulation study is described in Section 4. Application of the model to the ENABLE III RCT is provided in Section 5, followed by a discussion in Section 6.

## 2. MOTIVATING EXAMPLE

ENABLE III (Bakitas et al. 2015) was a two-arm palliative care RCT (conducted between October, 2009 and March, 2013) investigating the benefits of palliative care introduced soon after diagnosis (early group) compared with 12 weeks after randomization (delayed group) for patients with advanced cancer. Two hundred and seven patients were randomly allocated to either the early group ($n = 104$) or the delayed group ($n = 103$). QOL data assessed by the 46-item Functional Assessment of Chronic Illness Therapy-Palliative Care (FACIT-pal) were collected at baseline and at 6, 12, 18, and 24 weeks and every 12 weeks thereafter until death or study completion.

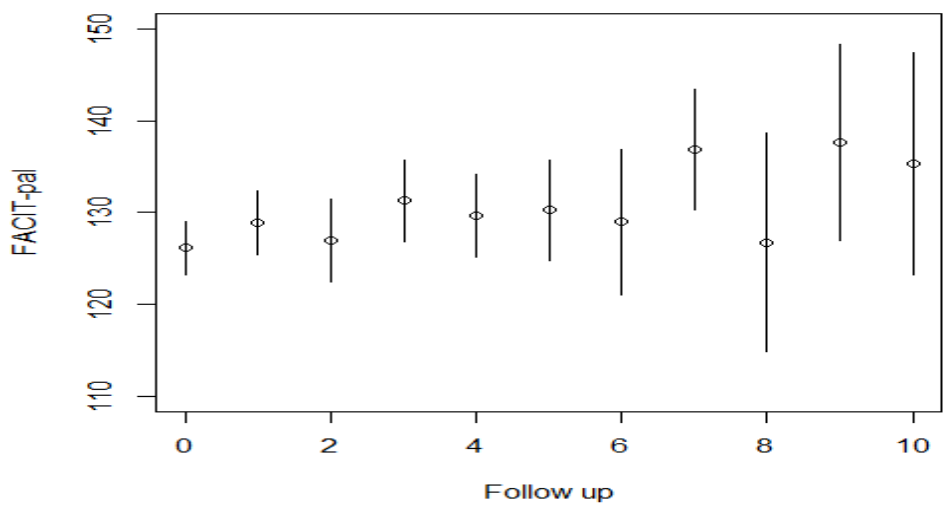

Figure 1. Mean QOL and the 95% confidence intervals (CI) for the first 10 follow up measurements where 0 denotes baseline.

Higher FACIT-pal scores indicate better QOL. Clinically, QOL is expected to decrease as a patient approaches death (Murray et al. 2005). However, in ENABLE III the overall pattern of observed mean QOL increased overtime (Figure 1). This paradoxical phenomenon can be explained by the survival distribution and the QOL terminal trend because patients who were very ill, and presumably a lower QOL, died sooner. Thus the mean QOL increased as the surviving patients were on average less ill. In ENABLE III, the hazard rate was higher in the first 12-month time period (results shown in Section 5), which is consistent with this explanation. When comparing

QOL between the two arms in a study such as ENABLE III, it is important to adjust for any survival differences because better QOL in the early treatment group could simply be due to patients' longer survival. Therefore, joint modeling of the terminal decline of QOL and survival data is essential to more accurately interpret the study's results.

## 3. METHOD

### 3.1 Notation and Model

Suppose there are $N$ patients in a study. Let $D_i$ denote the death time of the $i$th subject calculated as the duration from the date of enrollment to the date of death. Let $Y_i(t^*)$ denote the outcome, e.g. QOL, measured at time $t^*$ counting backward from the time of death for the $i$th patient. So the time $t^*$ is on a retrospective time scale starting from the death time. Let $A_i$ denote the binary treatment status with value of 1 being treatment and value of 0 being control for the $i$th patient, $i = 1, \ldots, N$. A time-varying treatment effect model for the mean trajectory of $Y_i(t^*)$ is given by

$$E[Y_i(t^*)|A_i] = \beta_\mu(t^*) + A_i \beta_A(t^*),$$

where $\beta_\mu(t^*)$ denotes the mean trajectory in the control group and the $\beta_A(t^*)$ denotes the time-varying treatment effect at time $t^*$ counting backward from the time of death for the $i$th patient. Both $\beta_\mu(t^*)$ and $\beta_A(t^*)$ are unspecified smooth functions. Under a mixed effects model framework for the longitudinal measurements (Diggle et al. 2002), the model adjusting for covariates is given by

$$Y_i(t^*) = \beta_\mu(t^*) + A_i \beta_A(t^*) + X_i^T \psi_X + Z_i^T b_i + \varepsilon_i(t^*), \tag{1}$$

where $\psi_X$ is the $P$-dimensional vector of fixed effect parameters associated with the covariate vector $X_i$ (the superscript $T$ denotes transpose of a vector or matrix) and $b_i$ are independent $L$-dimensional vectors of random effects associated with the covariates $Z_i$. In a randomized trial, no covariate adjustment is necessary for the standard intent-to-treat analysis, but is introduced here to cover observational studies as well. The random effects $b_i$ are assumed to follow a multivariate normal distribution $N(0, \Sigma)$ where $\Sigma$ is a $L$ by $L$ variance-covariance matrix. The residual error $\varepsilon_i(t^*)$ is assumed to be normally distributed with mean $E[\varepsilon_i(t^*)] = 0$ and $\text{var}[\varepsilon_i(t^*)]=\tau^2$, where $\varepsilon_i(t^*)$ and $\varepsilon_i(s^*)$ are independent for any $t^* \neq s^*$. The random effects $b_i$ and residual error $\varepsilon_i(t^*)$ are assumed independent as in Zhang et al. (1998).

Notice that time points of the longitudinal measurements on the retrospective time scale are unknown for those subjects whose death times are censored. Therefore, the retrospective time $t^*$ is missing because $D_i$ is missing. To resolve this issue, we model the survival time with a Cox model (Cox 1972) so that all those censored subjects can be included in the analysis. The survival model is on a prospective time scale and can be written as:

$$\lambda_i(t) = \lambda_o(t) \exp(A_i \alpha_A + \tilde{X}_i^T \alpha_X), \qquad (2)$$

where $t$ denote the time from enrollment on the prospective time scale and $\alpha_X$ denote the $Q$-dimensional parameters associated with the $Q$-dimensional covariates $\tilde{X}_i$. The death time may be right censored by another event time denoted by $C_i$ (e.g., withdrawal or study end). We assume that the censoring is independent of the longitudinal outcome and independent of the death time conditional on the covariates, which is true if there is only administrative censoring. The censoring times $C_i$ are also assumed to be independent of the random effects and the residual in model (1).

Models (1) and (2) form the two sub-models of our joint modeling approach for longitudinal and survival outcomes. Regression splines will be used to handle the unspecified terminal trend as detailed in the next section.

## 3.2 Estimation

Regression splines approach is proposed to estimate unspecified parameter functions $\beta_\mu(t^*)$ and $\beta_A(t^*)$ in sub-model (1). Natural cubic splines with B-splines basis (Eilers and Marx 1996, de Boor 2000, Wu and Zhang 2006) will be used with the knots equally spaced at the sample quantiles of the data. The procedure for determining the optimal number of knots is outlined in the next section. For the purpose of illustration and simplicity, we assume that the natural cubic splines for $\beta_\mu(t^*)$ and $\beta_A(t^*)$ have the same $k_1$ knots corresponding to $k_1$ basis functions. The exact same logic can be followed when they have different knots. Once all the knots have been specified, the longitudinal model can be converted into parametric form. Let $B_i^1, i = 1, \ldots, k_1$ denote the B-spline basis functions. Notice that the basis functions $B_i^1$ depend on $t^*$ and we suppressed $t^*$ in the notations for simplicity. By approximating the unspecified function with the basis functions: $\beta_\mu(t^*) = \sum_{k=1}^{k_1} \beta_k B_k^1$ and $\beta_A(t^*) = \sum_{k=1}^{k_1} \beta_{k+k_1} B_k^1$, sub-model (1) can be rewritten as

$$Y_i(t^*) = \mathcal{X}_i(t^*)^T \beta + Z_i^T b_i + \varepsilon_i(t^*) \tag{3}$$

where $\mathcal{X}_i(t^*)^T = \left(B_1^1, \ldots, B_{k_1}^1, A_i B_1^1, \ldots, A_i B_{k_1}^1, X_i^T\right)$, a $(2k_1 + P)$-dimensional vector of variables and $\beta = (\beta_1, \ldots, \beta_{k_1}, \beta_{k_1+1}, \ldots, \beta_{2k_1}, \psi_X)^T$. Maximum likelihood estimates (MLEs) will be obtained by maximizing the joint log-likelihood function for all parameters denoted by $\theta = (\alpha, \beta, \Sigma, \tau, \Lambda_0(t))$ where $\alpha = (\alpha_A, \alpha_X)$ and the cumulative baseline hazard function $\Lambda_0(t) = \int_0^t \lambda_o(s) ds$.

Let $Y_{ij}$ denote the outcome measured at $t_{ij}^*$ for the $i$th subject, $i = 1, \ldots, N; j = 1, \ldots, n_i$. Keep in mind that the time scale counts backward, which implies that $y_{i1}$ is the first measurement counting backward from the death time. The outcome vector $Y_i = (Y_{i1}, \ldots, Y_{in_i})^T$ has a multivariate normal distribution $N(\mu_i, V_i)$ with the mean vector $\mu_i$ having the $j$th element equal to $\mathcal{X}_i(t_{ij}^*)^T \beta$ and the variance matrix $V_i = \tau^2 I_{n_i} + Z_i \Sigma Z_i^T$, where $I_{n_i}$ is the $n_i$ by $n_i$ identity matrix.

Since censoring is assumed to be independent of the longitudinal outcome and the death time conditional on the covariates, the censoring times can then be treated as constants rather than random in the formulation of the likelihood function. We divide the sample into four groups based on two factors: censored or not and with or without follow up longitudinal measurements. The first group consists of the patients who had at least one follow-up measurement before the death was observed in the study. So their contribution to the log-likelihood function comes from both the longitudinal measurements and the observed death times. Without loss of generality, let $(i = 1, \ldots, r_1)$ denote patients in group 1. Their log-likelihood contribution is given by

$$l_1 = -0.5 \sum_{i=1}^{r_1} \left[ n_i \log(2\pi) + \log|V_i| + (Y_i - \mu_i)^T V_i^{-1} (Y_i - \mu_i) \right] + \sum_{i=1}^{r_1} \left[ \log(\lambda_o(D_i)) + A_i \alpha_A + \tilde{X}_i^T \alpha_X - \exp(A_i \alpha_A + \tilde{X}_i^T \alpha_X) \Lambda_0(D_i) \right].$$

The second group consists of those patients who had no longitudinal measurements before they died. Let $(i = r_1 + 1, \ldots, r_1 + r_2)$ denote the patients in group 2. Their log-likelihood contribution is given by the marginal likelihood of the observed survival times $l_2 = \sum_{i=r_1+1}^{r_1+r_2} \left[ \log(\lambda_o(D_i)) + A_i \alpha_A + \tilde{X}_i^T \alpha_X - \exp(A_i \alpha_A + \tilde{X}_i^T \alpha_X) \Lambda_0(D_i) \right].$

The third group consists of those patients who had at least one follow-up measurement and their death times were censored. Let $(i = r_1 + r_2 + 1, \ldots, r_1 + r_2 + r_3)$ denote the patients in group 3. Their contribution comes from the joint distribution of the longitudinal and survival data and the log-likelihood function is calculated as log of the integral of the conditional density of the longitudinal data over the survival distribution beyond the censoring time. It is given by

$$l_3 = \sum_{i=r_1+r_2+1}^{r_1+r_2+r_3} \log\left(\int_{C_i}^{+\infty} f(Y_i|D_i = s) \exp(A_i\alpha_A + \tilde{X}_i^T \alpha_X) \exp\left[-\exp(A_i\alpha_A + \tilde{X}_i^T \alpha_X)\Lambda_0(s)\right] d\Lambda_0(s)\right),$$

where $f(Y_i|D_i = s)$ is the conditional multivariate normal density function of the longitudinal outcome given death time is $s$. The joint likelihood function can be viewed as a weighted likelihood function of the longitudinal data with the weights being the density of the survival function beyond the censoring point.

The fourth group consists of those patients who had no longitudinal measurements before their death times were censored. Let $(i = r_1 + r_2 + r_3 + 1, \ldots, r_1 + r_2 + r_3 + r_4)$ denote the patients in group 4. Their log-likelihood contribution is given by the marginal likelihood of the censored survival times $l_4 = -\sum_{i=N-r_4+1}^{N} \exp(A_i\alpha_A + \tilde{X}_i^T \alpha_X)\Lambda_0(C_i)$.

Notice that the baseline hazard $\lambda_0(t)$ and the cumulative baseline hazard $\Lambda_0(t)$ are unspecified nonparametric curves in the model. We borrow the idea from the profile likelihood approach (Murphy and van der Vaart 2000) where $\Lambda_0(t)$ is profiled out by a step function, $\tilde{\Lambda}_0(t)$, starting at zero and jumping at observed death times. Let $d_1, d_2, \ldots, d_M$ denote the ascendingly ordered observed death times in the study. The $m$th jump at $d_m$ denoted by $\tilde{\Lambda}_0[d_m]$ is given by

$$\tilde{\Lambda}_0[d_m] = \frac{1}{\sum_{j \in R(d_m)} \exp(A_j\alpha_A + \tilde{X}_j^T \alpha_X)}, \quad m = 1, \ldots, M,$$

where $R(d_m)$ denotes the at-risk group at death time $d_m$. The step function $\tilde{\Lambda}_0(t)$ becomes the Breslow estimator of $\Lambda_0(t)$ when $\alpha_A$ and $\alpha_X$ are replaced by their MLE's. This approach essentially puts the survival distribution masses on the observed death times. We substitute $\tilde{\Lambda}_0(t)$ for $\Lambda_0(t)$ in $l_1, l_2, l_3$ and $l_4$ and substitute $\tilde{\Lambda}_0[D_i]$ for $\lambda_o(D_i)$ in $l_1$ and $l_2$ to obtain the new log-likelihood function. Let $l_1^*, l_2^*$ and $l_4^*$ denote the new log-likelihood functions corresponding to $l_1, l_2$ and $l_4$ after the substitution. It is straightforward to see that

$$l_1^* = -0.5 \sum_{i=1}^{r_1}\left[n_i \log(2\pi) + \log|V_i| + (Y_i - \mu_i)^T V_i^{-1}(Y_i - \mu_i)\right] + \sum_{i=1}^{r_1}\left[\log(\tilde{\Lambda}_0[D_i]) + A_i\alpha_A + \tilde{X}_i^T\alpha_X - \exp(A_i\alpha_A + \tilde{X}_i^T\alpha_X)\tilde{\Lambda}_0(D_i)\right],$$

$$l_2^* = \sum_{i=r_1+1}^{r_1+r_2}\left[\log(\tilde{\Lambda}_0[D_i]) + A_i\alpha_A + \tilde{X}_i^T\alpha_X - \exp(A_i\alpha_A + \tilde{X}_i^T\alpha_X)\tilde{\Lambda}_0(D_i)\right]$$

and

$$l_4^* = -\sum_{i=N-r_4+1}^{N} \exp(A_i\alpha_A + \tilde{X}_i^T\alpha_X)\tilde{\Lambda}_0(C_i).$$

The integral in $l_3$ becomes Stieltjes integral (Terhorst 1986) after the substitution. Let $l_3^*$ denote the new log-likelihood corresponding to $l_3$ after the substitution. It can be calculated as

$$l_3^* = \sum_{i=r_1+r_2+1}^{r_1+r_2+r_3} \log\left(\sum_{m=1}^{M} f(Y_i|D_i = d_m)1(d_m > C_i)P_{im}\right),$$

where $1(d_m > C_i)$ is an indicator function and $P_{im} = \exp[A_i\alpha_A + \tilde{X}_i^T\alpha_X - \exp(A_i\alpha_A + \tilde{X}_i^T\alpha_X)\tilde{\Lambda}_0(d_m)]\tilde{\Lambda}_0[d_m]$. Notice that $P_{im}$ is the probability of the $i$th subject dying at $d_m$, and thus $\sum_{m=1}^{M} 1(d_m > C_i)P_{im}$ should be the probability of surviving beyond $C_i$. However, $\sum_{m=1}^{M} 1(d_m > C_i)P_{im}$ is actually not equal to the survival function $P(D_i > C_i) = \exp\left(-\exp(A_i\alpha_A + \right.$

$\tilde{X}_i^T \alpha_X) \tilde{\Lambda}_0(C_i)\right)$ under the discretized survival distribution that put the masses on death times. This discrepancy could result in biased estimates. To correct this issue, we give a weight $W_i$ to $P_{im}$ so that $\sum_{m=1}^M 1(d_m > C_i) W_i P_{im} = \exp\left(-\exp(A_i \alpha_A + \tilde{X}_i^T \alpha_X) \tilde{\Lambda}_0(C_i)\right)$. The weight $W_i$ is then equal to $\frac{\exp\left(-\exp(A_i \alpha_A + \tilde{X}_i^T \alpha_X)\tilde{\Lambda}_0(C_i)\right)}{\sum_{m=1}^M 1(d_m > C_i) P_{im}}$. Let $P_{im}^* = W_i P_{im}$. Thus, we have the weighted $l_3^*$ given by

$$wl_3^* = \sum_{i=r_1+r_2+1}^{r_1+r_2+r_3} \log\left(\sum_{m=1}^M f(Y_i | D_i = d_m) 1(d_m > C_i) P_{im}^*\right).$$

By having the weight $W_i$ in $wl_3^*$, our approach will generate exactly the same estimates as the partial likelihood approach for $\alpha_A$ and $\alpha_X$ when the longitudinal and survival data are independent. Notice that the weight $W_i$ should converge to 1 in probability because both the numerator and denominator of $W_i$ converges to $\exp\left(-\exp(A_i \alpha_A + \tilde{X}_i^T \alpha_X) \Lambda_0(t)\right)$ as the Breslow estimator converges to $\Lambda_0(t)$ in probability (Zeng and Lin 2006). This means that the weight $W_i$ does not make a meaningful difference when there are enough number of subjects.

The complete log-likelihood function can then be calculated as

$$l = l_1^* + l_2^* + wl_3^* + l_4^*.$$

Parameter estimates are obtained by maximizing $l$ and variances are obtained from the Fisher information matrix. Point-wise CI can be constructed for the nonparametric curves $\beta_\mu(t^*)$ and $\beta_A(t^*)$ based on the estimated variance matrix of the parameters. This estimation approach puts the survival distribution masses on the observed death times which induces another potential issue for our joint model: the log-likelihood contribution is not well defined for those subjects in group 3 with death times censored beyond the last observed death time. To resolve this issue, we treat the last observed time as a death time regardless of its actual censoring status. This approach was studied by Efron (1967) who showed that the Kaplan-Meier estimator remains consistent. We will study the sensitivity of this assumption for our joint modeling approach in the simulation study.

**3.3 Knots Selection**

Regression splines approach has the advantage of treating the model as if it were a parametric model once the knots were determined, but it requires appropriately selecting the number and position of knots. We adopt the popular method placing the knots at equally spaced sample quantiles of the data. There have been quite a lot of procedures developed in the literature to determine the optimal number of knots including cross-validation (Hastie and Tibshirani 1990), generalized cross-validation (GCV) (Hastie and Tibshirani 1990, Nan et al. 2005), AIC and BIC. Generally, goodness of fit improves as a model becomes more complex (e.g., having more parameters). A good method for choosing knots is usually trying to select the number of knots such that it achieves the optimal balance between goodness of fit and model complexity. Goodness of fit can be measured by the sum of squared errors (SSE) or log-likelihood and the model complexity by the degrees of freedom (df) of the linear smoother or number of parameters that increase as the number of knots. For example, the GCV statistic is usually a ratio of an increasing function of (or weighted) SSE over a decreasing function of the df of the linear smoother, and thus, minimizing the GCV statistic results in a good balance between goodness of fit and model complexity. BIC and AIC are two other widely used model selection criteria that penalize the

-2*log-likelihood (goodness of fit) by the number of parameters (model complexity). For the estimation approach outlined in this paper, the BIC or AIC criteria are the most natural choices since the log-likelihood function has a closed form expression. Because the AIC criterion places a smaller penalty, relative to BIC, on model complexity, we have opted to use AIC to determine the optimal choice of knots. BIC gives similar results as shown in the simulation study. AIC statistic is defined by

$$AIC(k_1) = -2loglik + 2(\#parameters) = -2loglik + 2\left[2k_1 + P + \frac{L(L+1)}{2} + 1 + Q + 1\right]$$

The $k_1$ minimizing the AIC statistic will be chosen as the optimal number of knots. It is straightforward to extend the procedure to the case with $\beta_\mu(t^*)$ and $\beta_A(t^*)$ having different knots, but computational demand will increase as well on a higher order of magnitude.

### 3.4 Quality-Adjusted Life Years (QALY)

QALY is an important clinical endpoint and has been extensively discussed in the literature (Zhao and Tsiatis 1997, Cole et al. 1998, Cole et al. 2001, Cole et al. 2004, Wang and Zhao 2007, Jia and Lubetkin 2010). It is usually defined as a weighted life years in the form $\sum_{j=1}^{k} q_j S_j$ where $q_1, \dots, q_k$ are the discrete utility scores in each of the $k$ health states, and $S_1, \dots, S_k$ are measures of life duration spent in each state observed in a study. The definition of utility scores may vary from study to study, but it is usually standardized into the range from 0 to 1 with 0 indicating worst possible QOL and 1 best possible QOL. When the utility score is considered as a continuous variable denoted by $q(t)$, QALY becomes the area under the longitudinal curve of the utility score, i.e., $\int_0^{\Delta \wedge D} q(t) dt$, where $\Delta \wedge D = \min(\Delta, D)$, $D$ is the death time and $\Delta$ is the study duration or other time horizon for QALY. Under model (1) for the utility score and model (2) for the survival, the mean QALY can be calculated as

$$E\left[\int_0^{\Delta \wedge D} q(t)dt\right] = E\left[\int_0^{\Delta} q(t)1(t \leq D)dt\right] = \int_0^{\Delta} E[q(t)1(t \leq D)]dt$$

$$= \int_0^{\Delta} (\sum_{m=1}^{M} E[q(t)|d_m]1(t \leq d_m) \lambda(d_m)\exp(-\Lambda(d_m)))dt$$

$$= \int_0^{\Delta} \sum_{m=1}^{M} (\mathcal{X}_i(d_m - t)^T \beta) 1(t \leq d_m)\lambda(d_m) \exp(-\Lambda(d_m)) dt,$$

where $\lambda(.)$ and $\Lambda(.)$ denote the hazard and cumulative hazard functions respectively. By plugging the mean value of the covariates into the above formula, Least-Squares-mean (Gianola 1982) type of estimates of the QALY can then be obtained for both treatment and control groups. In the application of our model to ENABLE III discussed later, the continuous standardized QOL (original score divided by the maximum of the scale 184) is defined as the utility.

## 4. SIMULATION

Finite sample properties of the estimators were studied through a simulation study where QOL, survival data, and two covariates (baseline QOL and sex) were randomly generated for an RCT with 250 subjects in each arm. We generated 1000 samples where baseline QOL ($qol_0$) was generated with a Uniform distribution on (100, 150) and sex was generated with a Bernoulli distribution with the parameter $p = 0.5$. The models for the longitudinal QOL and survival outcome were as follows:

$$Y_i(t^*) = \beta_\mu(t^*) + A_i \beta_A(t^*) + \psi_1 qol_0 + \psi_2 sex + b_i + \epsilon_i(t^*)$$

$$\text{and } \lambda_i(t) = \lambda_o(t) \exp(A_i \alpha_A + \alpha_1 qol_0 + \alpha_2 sex),$$

where $b_i$ is the random intercept with the scalar variance $\sigma^2 = 36$ and the residual $\epsilon_i(t^*)$ has a scalar variance $\tau^2 = 16$. Here $\beta_\mu(t^*) = 140 - \frac{30}{1+0.2t}$, $\beta_A(t^*) = 30\exp(-0.23t - 0.92)$, $\psi_1 =$

0.3, $\psi_2 = 1$, $\lambda_o(t) = \exp(-2.5)$ and $\alpha_A = -0.2$, $\alpha_1 = -0.005$ and $\alpha_2 = 0.2$. The trajectory functions $\beta_\mu(t^*)$ and $\beta_A(t^*)$ were chosen to have similar patterns to what was found in Li et al (2013). Repeated measurements of QOL were assumed to be taken at every half month until death. Censoring was assumed to be independent of the longitudinal data and survival, and was generated by $G1(G < c) + c1(G \geq c)$ where $G$ follows a Gamma distribution and $c$ is constant. To make the censoring rates similar to ENABLE III where there was 47.34% total censoring, 36.71% censored subjects having longitudinal QOL measurements (i.e., in group 3) and 4.35% censoring beyond the last observed death time, we set the shape and scale parameters of $G$ equal to 1.55 and 14.3 respectively, $c = 33.65$, and the first measurement of QOL was set to be collected at $t = 5.5$. On average, there are 48.7% total censoring, 37.3% in group 3 and 5% censoring beyond the last observed death time in the simulated data sets. We selected the optimal number of knots for each sample within the range $k_1 \in [2, 11]$ using AIC. Results from the simulation showed that mean estimates of the natural cubic splines (dashed curves: Mean JM est.) were very close to the true function (the top panels of Figures 2 & 3). Results from a naïve approach were also presented in the figures for comparison which will be discussed later in this section. The pointwise coverage probabilities (CP) for 95% CI (dashed curves: JM 95% CP) were plotted in the bottom panels of Figures 2 & 3 with the mean coverage probabilities being 94.9% and 93.6% for $\beta_\mu(t^*)$ and $\beta_A(t^*)$ respectively. The average number of knots was 5.72 when AIC was used and 5.03 for BIC. The estimation results looked similar when BIC was used. The Wald-type 95% CI of estimators for the regression coefficient parameters $\psi_1, \psi_2, \alpha_A, \alpha_1$ and $\alpha_2$ (Table 1) had proper CP although there were some small biases for $\psi_1, \alpha_A, \alpha_1$ and a noticeable bias for $\psi_2$. The mean of the standard error estimators correctly reflected the empirical variability of the parameter estimators.

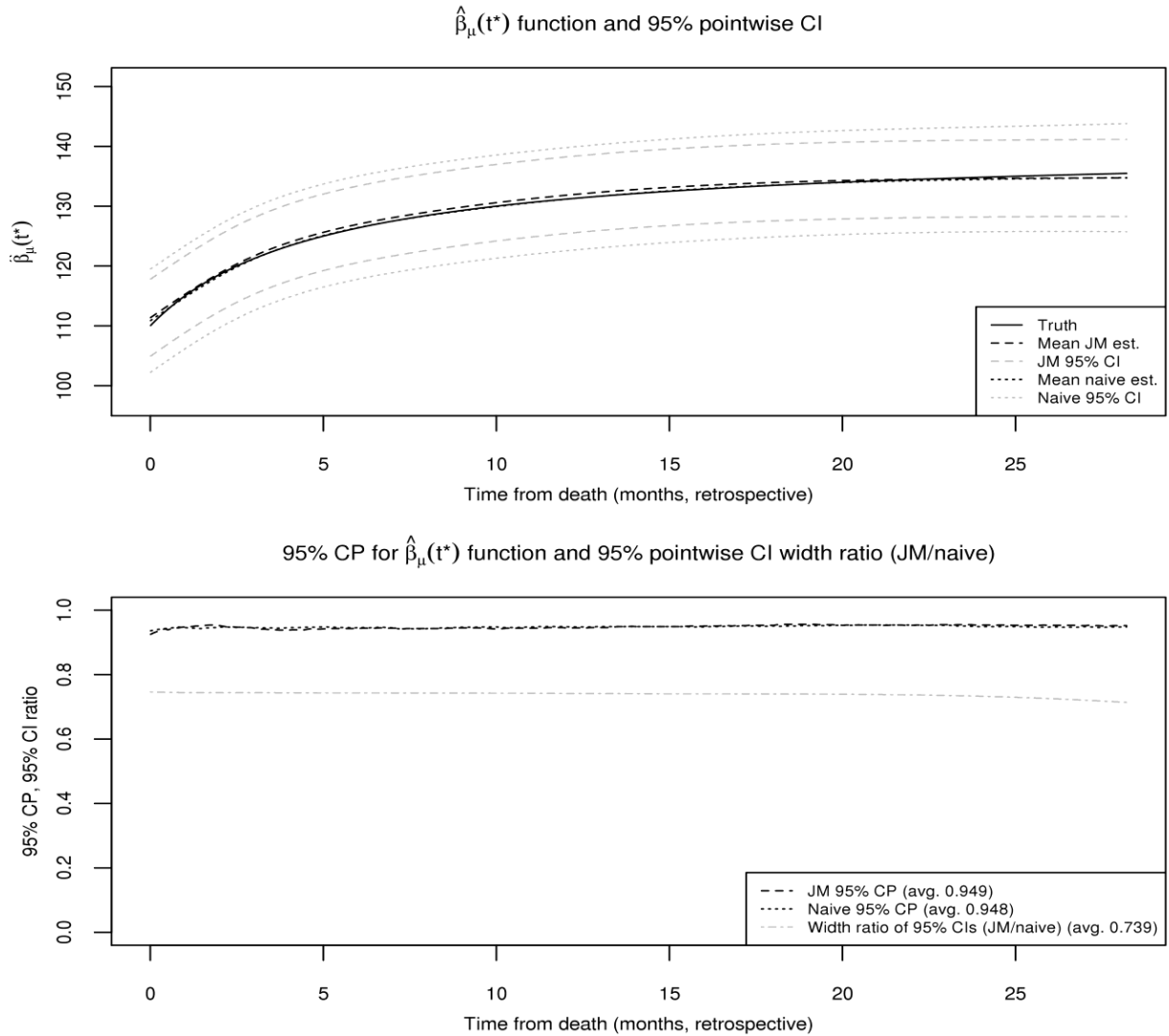

Figure 2. Estimation results for $\beta_\mu(t^*)$. The top panel shows the fit of the mean estiamted cuves from 1000 random samples along with the mean point-wise 95% CI for the joint modeling approach (JM) and the naïve approach. The bottom panel shows the point-wise 95% coverage probabilities for both methods and the ratio (JM/naive) of the width of the 95% CI.

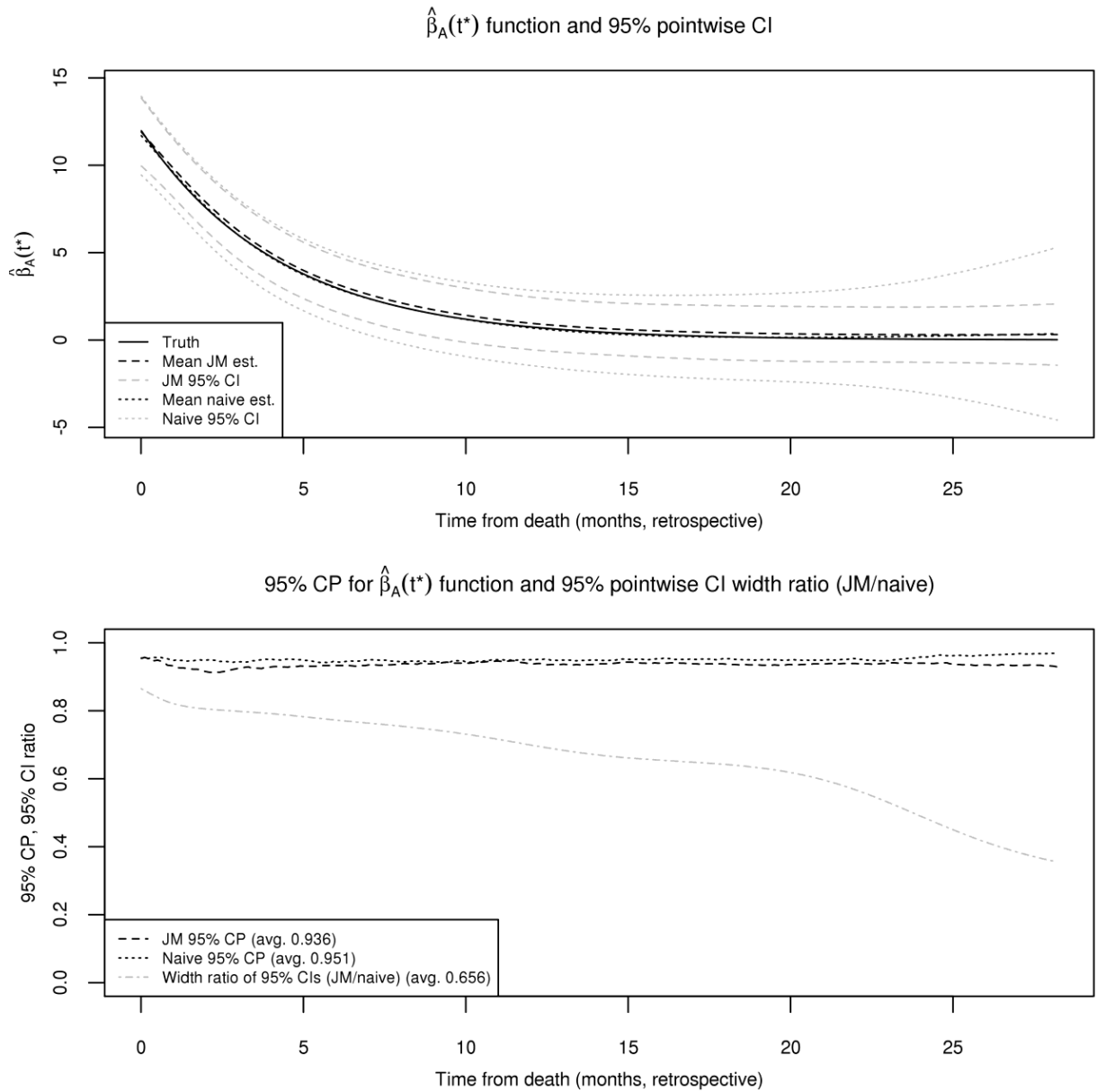

Figure 3. Estimation results for $\beta_A(t^*)$. The top panel shows the fit of the mean estimated cuves from 1000 random samples along with the mean point-wise 95% CI for the joint modeling approach (JM) and the naïve approach. The bottom panel shows the point-wise 95% coverage probabilities for both methods and the ratio (JM/naive) of the width of the 95% CI.

Table 1. Simulation results for the estimation of the parametric parameters. Bias is the mean of estimates minus the true values; Percent of Bias is the bias as a percentage of the true value. SE is the empirical standard error of the estimates; Mean SE is the mean of estimated standard errors; CP is the empirical coverage probability of the 95% CI for the parameters. Naïve Mean SE and Naïve CP are the counterparts from the naïve model.

| Longitudinal Model | | | | | | | | |
|---|---|---|---|---|---|---|---|---|
| Parameter | True | Bias | Percent of Bias (%) | SE | Mean.SE | Naïve Mean.SE | CP (%) | Naïve CP (%) |
| $\psi_1$ | 0.30 | 0.004 | 1.30 | 0.03 | 0.025 | 0.035 | 95.0 | 94.8 |
| $\psi_2$ | 1 | -0.27 | -27.29 | 0.68 | 0.71 | 1.00 | 94.9 | 94.8 |
| $\sigma$ | 6 | 0.21 | 3.53 | 0.26 | 0.26 | -- | 89.0 | -- |
| $\tau$ | 4 | <0.0001 | <0.0001 | 0.04 | 0.04 | -- | 93.6 | -- |
| Survival Model | | | | | | | | |
| $\alpha_A$ | -0.2 | -0.18 | -10.82 | 0.12 | 0.12 | 0.13 | 94.4 | 95.0 |
| $\alpha_1$ | -0.005 | -0.0001 | 2.13 | 0.004 | 0.004 | 0.004 | 94.3 | 93.7 |
| $\alpha_2$ | 0.2 | 0.002 | 1.25 | 0.12 | 0.12 | 0.13 | 95.1 | 94.4 |

We compared our joint modeling method with the naïve method (dotted curves in Figures 2 & 3) where censored subjects were simply omitted for the longitudinal model (1) although the survival model (2) in the naïve analysis used all subjects. The naïve method generated less bias (results not shown), but our joint modeling method was much more efficient. The joint model approach achieved comparable CP with 26.1% narrower CI on average for $\beta_\mu(t^*)$ and 34.4% narrower for $\beta_A(t^*)$. In other words, the standard error (SE) estimates from the joint model were on average 26.1% and 34.4% smaller for $\beta_\mu(t^*)$ and $\beta_A(t^*)$ respectively. The point-wise SE at large time points for $\beta_A(t^*)$ could be over 60% smaller as indicated by the dash-dotted curve in the bottom panel of Figure 3. For the regression coefficient parameters in the model (Table 1), our approach achieves slightly better CP with 26.2% and 28.4% smaller SE estimates for $\psi_1$ and $\psi_2$ respectively, and 2.4%, 3.1% and 3.1% smaller SE estimates for $\alpha_A$, $\alpha_1$ and $\alpha_2$ respectively. Therefore our joint modeling method was much more efficient than the naïve approach for estimating the longitudinal parameters. The main reason for the efficiency gain was that our method used all the subjects in the modeling whereas the naïve method omitted the longitudinal QOL data from the censored subjects. Incorporating the association between the longitudinal QOL and survival outcomes also increased the efficiency reflected in the smaller SE estimates for the survival parameters. Even when the longitudinal and survival outcomes were independent, we still observed significant efficiency gain for estimating the longitudinal parameters (results not shown) although the survival parameter estimates were exactly the same as the naïve analysis.

The overall censoring rate and the size of group 3 (defined in the Section 4 for estimation) in the ENABLE III study is on the high end relative to other palliative care studies including ENABLE II (Bakitas et al. 2009) and Temel et al. (2010). ENABLE II had 28.26% overall censoring with 18.94% in group 3. Simulation results (not shown) looked similar under the ENABLE II censoring rates and another intermediate censoring rates between ENABLE II and III (37% overall censoring and 27% in group 3). We also carried out a sensitivity study to examine the assumption that the last observed time is assumed to be a death time regardless of its actual censoring status. In both ENABLE II and III studies, the rates of censoring beyond the last observed death time (CBLD) happened to be the same and equal to 4.35%. It is unlikely to see a very high CBLD rate in palliative care studies since all patients have a relatively short prognosis. In the simulation settings described above, the average CBLD rates were set at 5%. Starting with the above setting with 48.7% total censoring, 37.3% in group 3 and 5% CBLD, we increased CBLD rate by lowering $c$, the upper bound of the censoring variable, to get 10%, 12.5% and 15% CBLD. We found that the biases increased with the CBLD rate for both parametric and nonparametric parameters and some CP fell below 90% for 12.5% and 15% CBLD although the CP for the parametric regression coefficients were relatively more stable.

## 5. APPLICATION

We applied our model to the recently completed ENABLE III RCT (Bakitas et al. 2015) described in Section 2. In ENABLE III there were 103 subjects in the early treatment group and 104 subjects in the delayed treatment group. Overall, the censoring rate was 47.34%; 36.71% censored subjects had longitudinal QOL measurements and 4.35% CBLD. The baseline QOL ($QoL_0$) and sex variables were adjusted in both the sub-models given by

$$Y_i(t^*) = \beta_\mu(t^*) + A_i\beta_A(t^*) + \psi_1 QoL_0 + \psi_2 \text{sex} + b_i + \epsilon_i(t^*), \tag{5}$$

$$\text{and } \lambda_i(t) = \lambda_o(t) \exp[A_i\alpha_A + \alpha_1 QoL_0 + \alpha_2 \text{sex}]. \tag{6}$$

We used the same knots for the natural cubic splines estimating $\beta_\mu(t^*)$ and $\beta_A(t^*)$ in sub-model (5). Based on the AIC criterion, the optimal number of knots was 7. The fitted longitudinal trajectories using natural cubic splines in model (5) are shown in Figure 4 with 95% pointwise CI. Other parameters are given in Table 2. The slope of cumulative hazard function (Figure 5) was slightly larger until approximately 12 months for both groups indicating that the hazard of death was higher during that period. This explains the increasing pattern in Figure 1 since patients who survived beyond 12 months were likely less ill and presumably had better QOL. Only the baseline QOL (associated parameters are $\psi_1$ and $\alpha_1$) was a significant predictor in both sub-models. The findings for the treatment effects were null at all time points which is consistent with the results presented in Bakitas et al. (2015). We also analyzed the data using naïve approach (results not shown) and the efficiency gain of our joint modeling approach for the longitudinal parameters was obvious. We also estimated the terminal trend using the piecewise linear joint modeling approach proposed in Li et al. (2013) and the results looked similar (Figure 4). Estimates and comparisons for QALY are presented in Table 3 for two time horizons: 33.5 months, the maximum observed time, and 12 months, to compare earlier time horizons. The mean QALY and 95% CI are given in Table 3 where the delta method was used to obtain standard errors.

| Table 2. Estimates of model parameters | | | |
|---|---|---|---|
| Longitudinal Model | | Survival Model | |
| Parameter | Estimate (SE) | Parameter | Estimate (SE) |
| $\psi_1$ | 0.59 (0.05) | $\alpha_A$ | -0.31 (0.19) |
| $\psi_2$ | 2.58 (2.23) | $\alpha_1$ | -0.01 (0.005) |
| $\sigma$ | 11.57 (0.99) | $\alpha_2$ | 0.34 (0.20) |
| $\tau$ | 11.93 (0.41) | | |

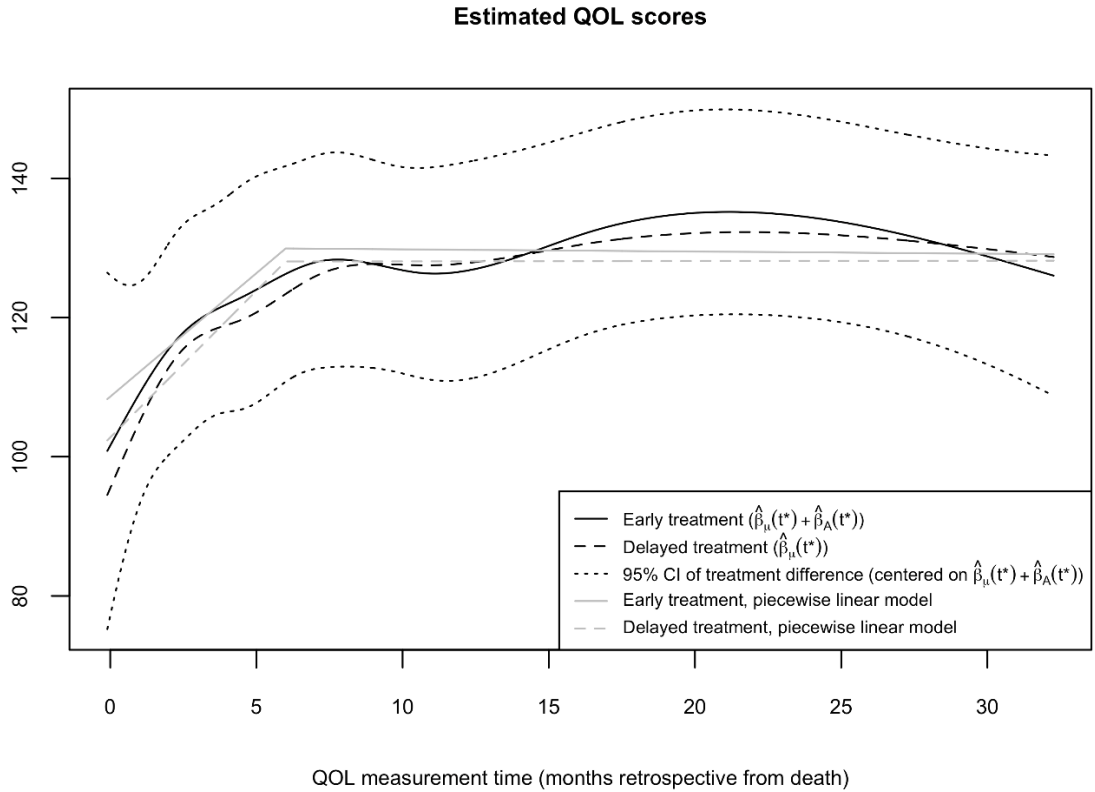

Figure 4. Fitted longitudinal trajectories for the early and delayed treatment groups and the point-wise 95% CI using the ENABLE III data. The piecewise linear curves (in grey) were obtained using the method in Li et al. (2013)

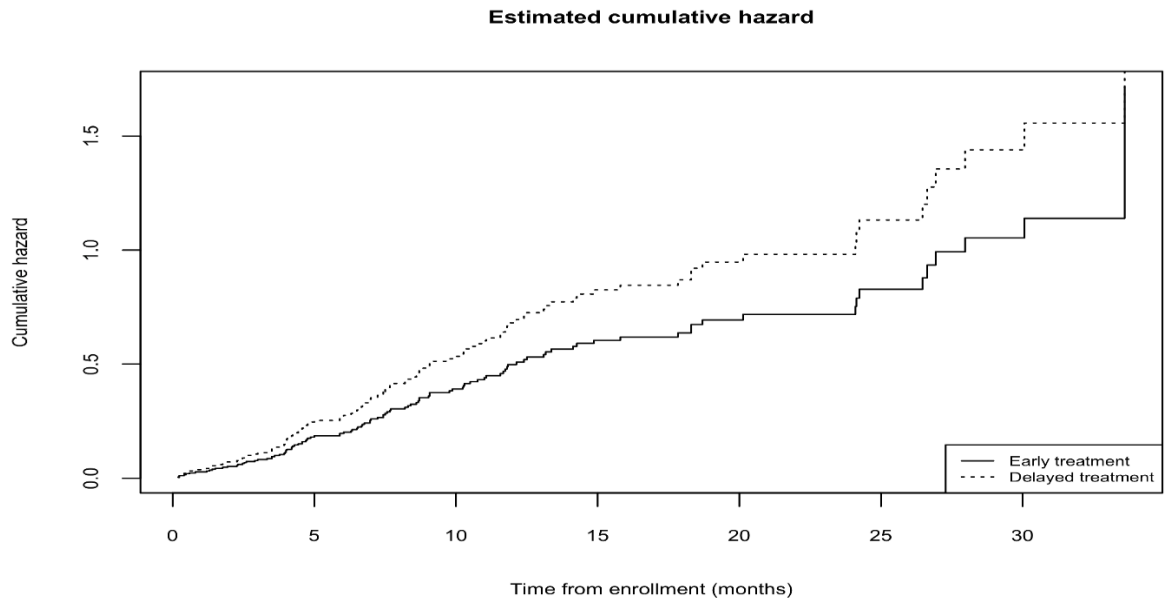

Figure 5. Fitted cumulative hazard functions at the mean values of the adjusted covariates for the early treatment group and delayed treatment groups using the ENABLE III data.

| Table 3. Estimated mean QALY with 95% CI in ENABLE III | | |
|---|---|---|
|  | Early treatment group | Delayed treatment group |
| Mean QALY in months | 33.5 months: 8.45 (8.04, 8.86) | 33.5 months: 8.34 (7.94, 8.74) |
|  | 12 months: 4.96 (4.76, 5.16) | 12 months: 4.90 (4.71, 5.08) |

# 6. DISCUSSION

With the intention to establish a cutting-edge statistical approach for analyzing longitudinal QOL and survival data in palliative care research, we developed a novel semiparametric joint modeling method that directly estimates terminal trends of longitudinal QOL outcomes while incorporating censored survival information. The terminal trend estimation can help understand how patients may die in the (relatively) near future which is the key to caring well for patients and then plan appropriately (Murray et al. 2005). Our approach can be applied to a variety of longitudinal outcomes and EOL research studies as well although the focus in this paper is longitudinal QOL outcome in palliative care studies. The nonparametric trajectory of the longitudinal QOL outcomes in the model makes the model highly flexible and not subject to model misspecification, and thus the model can be applied to other types of diseases as well as cancer. This approach appropriately accounts for the association between survival and longitudinal QOL outcomes, thus providing a more clinically meaningful estimate of the palliative care treatment effect. It accomplishes this by modeling the trajectories of the longitudinal QOL outcomes on a retrospective time scale. Dependence between censored death times (unknown time origins on a retrospective time scale) and the longitudinal QOL outcomes was appropriately modeled in the approach by constructing the joint likelihood function. One can also make inferences about QALY using this novel joint modeling approach. An R program to run the model is available upon request.

Compared with the fully parametric approach developed in Li et al. (2013) for studies with cancer patients where they require pre-specification of breakpoints for the piecewise models, this semiparametric model has more flexible nonparametric mean structure of the longitudinal trajectory that is not subject to model misspecification. This advantage enables the semiparametric approach to model the longitudinal terminal trends for studies with non-cancer patients. Compared with other joint models briefly reviewed in the introduction, our approach directly models the longitudinal QOL outcomes prior to death on a retrospective time scale. As a result, interpretation of such an analysis in palliative care research is likely more reflective of the nature to study purposes. Our joint model is more efficient (generating smaller SE) than a naïve analysis that uses the decedents subsample for the longitudinal model and the full sample for the survival model as shown in the simulation study; although the efficiency gain for the survival model is small (3.1%).

A few possible extensions of our model are worth mentioning. Besides regression splines that were employed in the model, other nonparametric approaches (e.g., penalized splines, smoothing splines) could also be used and the estimation procedure will need to be modified accordingly. In the current approach, only the intercept and treatment effect are modeled with nonparametric functions in the longitudinal model. It can be easily extended to a more general case where the coefficients of other covariates are nonparametric functions as well. Another possible extension of the model is to add a shared random effect in the survival model (Liu, Wolfe, and Kalbfleisch 2007) to account for another level of association between survival and the longitudinal QOL outcomes. To the best of our knowledge, there are limited developments for model diagnostic techniques for joint modeling in the literature. For shared parameter models, some informal diagnostic graphical procedures and multiple-imputation-based approaches to construct residuals and diagnostic plots can be found in the literature (Dobson and Henderson 2003, Rizopoulos, Verbeke, and Molenberghs 2010). Developments for model diagnostic tools under the terminal trend model may warrant further investigation.

Comorbidities could be a potential issue because terminal trajectories are usually different for different types of illnesses. Patients with multiple disorders may have multiple trajectories running concurrently, with the more rapidly progressing trajectory typically dominating others (Murray et al. 2005). This is not uncommon in older patients with slowly progressive cancers and thus more complicated models may be needed for this case. Comorbidities could also be related to another issue as pointed out by Currow et al. (2012) that in palliative care RCTs, subjects often die as a direct result of disease progression or drop out for reasons unrelated to the treatment. Improper handling of those patients could make estimates systematically biased away from the true effect. They proposed a modified intent-to-treat analysis approach to address this issue by identifying those who withdraw or die unrelated to the treatment and dropping them from the analysis. An alternative and better way to address this issue might be a joint modeling approach that appropriately models the survival and censoring distribution in terms of the treatment status to account for the informative dropouts.

Longitudinal QOL data missing at random is an assumption that is oftentimes unverifiable in practice. In the ENABLE III RCT, longitudinal QOL measurements were missing with an intermittent pattern and the reasons include "too busy", "lost interest in the study", "too ill to respond" and "unknown". Missing data as a result of being "too ill" might indicate that the missing longitudinal data was not missing at random; in such cases missing data could result in a bias in estimates. The semiparametric mixed effects model in the sub-model (1) for longitudinal QOL measurements should account well for the missing at random assumption since that is one of the advantages of mixed effects models. However, the results will be biased if data is not missing at random (or non-ignorable) and incorporating additional assumptions for the missing mechanism into the model could be helpful (Daniels and Hogan 2008).

In this paper, the censoring distribution is assumed to be independent of the survival and the longitudinal outcomes. In general, this might not be the case although it is true in ENABLE III since survival status of all patients was identified at the end of ENABLE III trial, and hence there was only administrative censoring, a form of non-informative censoring. Under this assumption, efficiency is the only advantage of our joint modeling compared with the naïve analysis approach. However, if the censoring is informative, for example, censoring distribution depends on QOL, the naïve analysis will generate biased results because the decedent subsample might be a relatively less ill subsample than others. In this case, our joint modeling approach can be modified to generate unbiased results by incorporating the association of the censoring distribution with QOL in the model.

**Acknowledgement**

Zhigang Li, H. R. Frost and Tor D. Tosteson are partially supported by R03NR014915. Zhigang Li is also partially supported by P01ES022832 and P20GM104416. H. R. Frost is also partially supported by P20GM103534. Tor D. Tosteson is also partially supported by the Dartmouth Clinical and Translational Science Institute, under award number UL1TR001086, and award number P30CA023108 to the Norris Cotton Cancer Center. Lei Liu is partially supported by AHRQ R01HS020263. Kathleen D. Lyons is supported by a Mentored Research Scholar Grant in Applied and Clinical Research (MRSG 12-113-01 – CPPB) from the American Cancer Society. Marie Bakitas is supported by the National Institute for Nursing Research (R01NR011871), National Cancer Institute (R01 CA101704), Foundation for Informed Medical Decision-Making and a Cancer and Leukemia Group B Foundation Clinical Scholar Award.

**Conflict of Interest**: None declared